\documentstyle[12pt]{article}
\newlength{\mathspace}
\def\np#1{ Nucl. Phys. B#1}
\def\pr#1    { Phys. Rev. D#1 }
\def\pl#1{ Phys. Lett. B#1}
\def\cmp   { Commun. Math. Phys. }

\def\ijmp#1  { Int. Jour. Mod. Phys. A#1 }
\def\mpl#1   { Mod. Phys. Lett. A#1 }
\def\begineq{\begin{equation}}
\def\endeq{\end{equation}}
\def\eqabegin{\begin{eqnarray}}
\def\eqaend{\end{eqnarray}}
\def\nn{\nonumber}

\def\parbigskip        {  \par\bigskip  }
\def\parmedskip        {  \par\medskip  }

\def\parbigskipn        {  \par\bigskip\noindent  }

\begin{document}
\baselineskip=0.7cm
\setlength{\mathspace}{2.5mm}




\begin{titlepage}

    \begin{normalsize}
     \begin{flushright}
                 UG-5/95, US-FT-2/95, ENSLAPP-L-504/95 \\
                 hep-th/9502090\\
     \end{flushright}
    \end{normalsize}
    \begin{large}
       \begin{center}
          {\bf  $N=4$: A Unifying Framework for\\
           $2d$ Topological Gravity, $c_M \leq 1$ String Theory and \\
           Constrained Topological Sigma Model\\}
       \end{center}
    \end{large}
\vspace{2mm}
\begin{center}
           Pablo M. L{\sc latas},
           \footnote{E-mail address:
              llatas@th.rug.nl}
           \footnote{Address after April 1, 1995: Department of Physics,
                    University of California at Santa Barbara, CA 93106, USA.}
            Shibaji R{\sc oy}
           \footnote{E-mail address:
              roy@th.rug.nl}
           \footnote{Address after February 15, 1995: Departamento de
                     F\'\i sica
                     de Particulas, Universidade de Santiago, E-15706,
                     Santiago de Compostela, Spain.}\\
      \vspace{2mm}
        {\it Institute for Theoretical Physics} \\
        {\it Nijenborgh 4, 9747 AG Groningen}, {\it The Netherlands}\\
     \vspace{2mm}
        and\\
     \vspace{2mm}
        J. M. S{\sc anchez de} S{\sc antos}
        \footnote{E-mail address: santos@gaes.usc.es}\\
\vspace{2mm}
        {\it Departamento de F\'\i sica de Particulas},
        {\it Universidade de Santiago}\\
        {\it E-15706 Santiago de Compostela, Spain}\\
        {\it and}\\
        {\it Laboratoire de Physique Th\'eorique} \footnote{URA 1436 du CNRS,
associ\'ee
 a l'ENS Lyon}, {\it ENS Lyon}\\
        {\it 46, all\'ee d'Italie, 69364 Lyon Cedex 07, France}
      \vspace{.5cm}

    \begin{large} {\bf Abstract} \end{large}
        \par
\end{center}

 \begin{normalsize}
\ \ \ \
It is shown that two dimensional ($2d$) topological gravity in the conformal
gauge has a larger symmetry than has been hitherto recognized; in the
formulation of Labastida, Pernici and Witten it contains a twisted ``small''
$N=4$ superconformal symmetry. There are in fact two distinct twisted $N=2$
structures within this $N=4$, one of which is shown to be isomorphic to
the algebra discussed by the Verlindes and the other corresponds, through
bosonization, to $c_M\leq 1$ string theory discussed by Bershadsky et.al.
As a byproduct, we find a twisted $N=4$ structure in $c_M\leq 1$ string
theory. We also study the ``mirror'' of this twisted $N=4$ algebra and find
that it corresponds, through another bosonization, to a constrained
topological sigma model in complex dimension one.
 \end{normalsize}

\end{titlepage}
\vfil\eject

Our understanding of the underlying symmetry structure of string theory
will be greatly facilitated in its formulation in a phase where the basic
symmetries are unbroken. It has been suggested that topological gravity [1]
and topological sigma models [2] provide such a framework namely, they
describe
the unbroken phase of ordinary gravity and the usual sigma models. This
exotic gravitational theory where the general covariance remains unbroken
has been formulated and clarified in two space--time dimensions by several
groups [3--5]. In the original formulation of $2d$ topological gravity
Labastida,
Pernici and Witten started with a Lagrangian which is identically zero and
therefore contains all the symmetries of the world.
The basic field is the $2d$ metric, and the symmetries are chosen
to be the combined diffeomorphism and an arbitrary shift of the metric
denoted by a symmetric local GL(2,R) parameter. These symmetries are then
fixed by introducing ghosts and ghosts of ghosts to obtain topological
gravity action as a free conformal field theory.

In a somewhat different formulation Verlinde and Verlinde [6] wrote down
$2d$
topological gravity action as a topological gauge theory where the gauge
group is the $2d$ Poincare group ISO(2). In their formulation supersymmetry
plays a more fundamental role. It has also been shown in [7] that $2d$
topological
gravity in the conformal gauge possesses a large class of symmetries where
the associated symmetry generators satisfy a twisted $N=2$ superconformal
algebra. The associated superconformal model has central charge $c^{N=2}
= -9$ indicating that the superconformal model is non-unitary.

In this paper, we consider the topological gravity action of Labastida,
Pernici and Witten (LPW). In the conformal gauge, we show that this action
has
a twisted $N=2$ superconformal symmetry which is isomorphic to the symmetry
algebra discussed by Verlinde and Verlinde. We also show, that the topological
gravity action of LPW contains another distinct twisted $N=2$ superconformal
symmetry where
the topological charge is the BRST charge associated with the diffeomorphism
invariance. These two twisted $N=2$ superconformal algebras are then shown to
combine to form a bigger algebra ----- a
 twisted ``small" $N=4$ superconformal algebra [8,9] with the associated
central charge $-$9. The topological gravity action, in fact, is invariant
under these twisted $N=4$ symmetry charges. In order to find an
interpretation of the latter twisted $N=2$ structure we make use of a
bosonization [10] by which we relate the fields in the topological gravity
to those of $c_M\leq 1$ string theory. We find that this $N=2$ structure is
precisely the same as appeared in the context of $c_M\leq 1$ string theory
in ref.[11--13] with a particular central charge. In this way we also recover
the full twisted ``small'' $N=4$ structure in $c_M\leq 1$ string theory as
well. We then study the ``mirror'' [14] of this twisted $N=4$ algebra. We
find that by properly identifying the fields and by making use of another
bosonization [15] they represent the generators of a constrained topological
sigma model in complex dimension one. Thus, twisted $N=4$ superconformal
algebra provides a unifying framework for 2d topological gravity,
$c_M\leq 1$ string theory and constrained topological sigma model.

The gauge fixed action of $2d$ topological gravity formulated
in [3] has the form,
\begineq
S = 2 \int_{\Sigma} d^2\sigma \sqrt{-g}\Big(-i b^{ij} D_i c_j
- B^{ij} D_i \phi_j\Big)
\endeq
Here integration is over the $2d$ compact, boundaryless manifold $\Sigma$.
$i, j$ are the $2d$ indices, $g_{ij}$ is a fixed metric on $\Sigma$ and
covariant derivatives are with respect to this metric. $(b^{ij}, c^i)$ is
the fermionic reparametrization ghost system whereas $(B^{ij}, \phi^i)$ is
the bosonic deformation ghost of ghost system. The fields $b^{ij}$ and
$B^{ij}$ are symmetric and traceless. The above action is invariant under the
following BRST transformations,
\eqabegin
\delta_Q b^{ij} &=& \epsilon_Q \Big(i b^{k(i} D_k c^{j)} - 2 i b^{ij} D_k c^k
                    -i (D_k b^{ij}) c^k
                 + B^{k(i} D_k \phi^{j)} - 2 B^{ij} D_k \phi^k\Big.\nn\\
                  & &\qquad \Big. - (D_k B^{ij})\phi^k
                 -i g^{ij} b^{kl} D_k c_l - g^{ij} B^{kl}
                      D_k \phi_l\Big)\\
\delta_Q c^i &=& \epsilon_Q \Big(i c^k D_k c^i + \phi^i\Big)\\
\delta_Q B^{ij} &=& i \epsilon_Q \Big( 2 B^{ij} D_k c^k - B^{k(i} D_k c^{j)}
                    + (D_k B^{ij}) c^k
                 +g^{ij} B^{kl} D_k c_l +b^{ij}\Big)\\
\delta_Q \phi^i &=& i \epsilon_Q \Big( c^k D_k \phi^i - (D_k c^i)\phi^k\Big)
\eqaend
where $\epsilon_Q$ is an infinitesimal fermionic parameter and we have defined
$A^{(i} B^{j)} \equiv A^i B^j + A^j B^i$. It is also straightforward to check
from (1) that the right hand side of (2) is the energy--momentum tensor
defined as,
\begineq
T^{ij} = -{1\over \sqrt {-g}} { \delta S \over \delta g_{ij}}
\endeq
Therefore, Eq.(2) tells us that the energy--momentum tensor is BRST-exact, one
of the basic properties of a topological field theory [16]. Since (1) defines a
free conformal system we rewrite the action in the conformal gauge
$g_{z{\bar z}} = g_{{\bar z} z} = \frac {1}{2}$, $ g_{zz} = g_{{\bar z}
{\bar z}}
= 0$. With the redefinitions $b_{zz} \equiv b$, $c^z \equiv -i c$,
$B_{zz} \equiv
-i\beta$, $\phi^z \equiv -i\gamma$ and similarly $b_{{\bar z} {\bar z}}\equiv
{\bar b}$, $c^{{\bar z}} \equiv - i {\bar c}$, $B_{{\bar z}{\bar z}}
\equiv -i{\bar
\beta}$, $\phi^{{\bar z}}\equiv -i{\bar \gamma}$, the action (1) takes the more
familiar form [17],
\begineq
S = \int_\Sigma d^2 z \big(-b {\bar \partial}c
+ \beta {\bar \partial}\gamma + {\rm h.c.}\big)
\endeq
The BRST transformations (2--5) in this gauge reduces to,
\eqabegin
\delta_Q b(z) &=& \epsilon_Q T(z)\\
\delta_Q c(z) &=& \epsilon_Q \Big( c(z)\partial c(z) + \gamma(z)\Big)\\
\delta_Q \beta(z) &=& \epsilon_Q \Big( G(z)- 2b(z)\Big)\\
\delta_Q \gamma(z) &=& \epsilon_Q \Big( c(z) \partial\gamma(z) -\partial
c(z) \gamma(z)\Big)
\eqaend
and similar expressions for the fields ${\bar b({\bar z})}, {\bar c({\bar z})},
{\bar \beta ({\bar z})}$ and $ {\bar \gamma({\bar z})}$. Since here we are
dealing with a free conformal field theory, the fields split up into
holomorphic and antiholomorphic sectors. We only concentrate on the holomorphic
sector. In (8) $T(z)$ is given by,
\begineq
T(z) = 2 :\beta(z)\partial\gamma(z): + :\partial \beta(z)\gamma(z):
        -2:b(z)\partial c(z): - :\partial b(z) c(z):
\endeq
and also $G(z)$ in (10) is defined as
\begineq
G(z) = G_s(z) + G_v(z)
\endeq
where,
\eqabegin
G_s(z) &=& 2 \beta(z)\partial c(z) + \partial \beta(z) c(z)\nn\\
G_v(z) &=& b(z)
\eqaend
Here $T(z)$ is the energy-momentum tensor and $G(z)$ its $Q$-partner
in the conformal gauge. It has been noted by Verlinde and Verlinde that
$2d$ topological gravity in the conformal gauge possesses a large class
of symmetries [6,7]. In fact, all known topological conformal field
theories contain a twisted $N=2$ superconformal algebra (TCA) generated
by two bosonic $\Big(T(z)\, {\rm and}\, J(z)\Big)$ and two fermionic
$\Big(G(z)\,{\rm and}\, Q(z)\Big)$
currents. The operator product expansions (OPE) among the currents are
given by:
\eqabegin
T(z)T(w)&\sim&\frac{2T(w)}{(z-w)^2} +\frac{\partial T(w)}{(z-w)}\\
T(z)Q(w)&\sim&\frac{Q(w)}{(z-w)^2} +\frac{\partial Q(w)}{(z-w)}\\
T(z)G(w)&\sim&\frac{2G(w)}{(z-w)^2} +\frac{\partial G(w)}{(z-w)}\\
T(z)J(w)&\sim& -\frac{c^{N=2}/3}{(z-w)^3} +\frac{J(w)}{(z-w)^2} +
 \frac{\partial J(w)}{(z-w)}\\
J(z)Q(w)&\sim&\frac{Q(w)}{(z-w)}\\
J(z)G(w)&\sim& -\frac{G(w)}{(z-w)}\\
J(z)J(w)&\sim&\frac{c^{N=2}/3}{(z-w)^2}\\
Q(z)G(w)&\sim&\frac{c^{N=2}/3}{(z-w)^3} +\frac{J(w)}{(z-w)^2} +
 \frac{T(w)}{(z-w)}\\
Q(z)Q(w)&\sim& G(z)G(w)\sim 0
\eqaend

The fermionic currents $Q(z)$ and $G(z)$ have conformal weights 1 and 2
respectively. Also, the $U(1)$ current $J(z)$ is a quasi-primary field.
We observe as well that the $U(1)$ charges of $Q(z)$ and $G(z)$ are $+1$ and
$-1$ respectively. Finally, we note from (22) that the
energy-momentum tensor is in fact $Q$-exact. In order to find this
structure in the topological gravity formulation of LPW, we first use the
$Q$-transformations
given in (8--11) and then derive the corresponding current by
 Noether's method. The resulting current is:
\eqabegin
\hat{Q} (z)&=&2c(z):\beta (z)\partial\gamma (z):+c(z):\partial\beta (z)
\gamma (z):-:c(z)b(z)\partial c(z):\nn\\
& &\qquad\qquad -\partial \Big(:\beta (z)\gamma (z):c(z)
\Big) +b(z)\gamma (z).
\eqaend
It is easy to verify that this current $\hat{Q} (z)$ is not a good
candidate for the twisted $N=2$ generator $Q(z)$, since the OPE of
$\hat{Q} (z)$ with itself is not regular (note (23)), but given as:\footnote{
The basic OPE's we use are $\beta (z)\gamma (w)\sim\frac{1}{z-w}$ and
$b(z)c(w)\sim\frac{1}{z-w}$.}
\eqabegin
\hat{Q} (z)\hat{Q} (w)\sim\frac{2\gamma (w)}{(z-w)^3}+\frac{\partial\gamma
(w)}{(z-w)^2}+\frac{\partial \Big(\gamma (w):c(w)b(w):+\gamma^2 (w)b(w)
+\frac{3}{2}\partial\gamma (w)\Big)}{(z-w)}.
\eqaend

We observe that it is possible to modify $\hat{Q} (z)$ by adding total
derivative terms such that the OPE becomes regular. The new current
in that case takes the form:
\eqabegin
Q(z)=\hat{Q} (z)-\partial \Big(:\beta (z)\gamma (z):c(z)\Big)-
\frac{3}{2}\partial^2
c(z).
\eqaend
Note that with this modification the basic relations (8--11) remain untouched.

It is also clear from (10), (13) and (14) that, since the BRST charge is
nilpotent, we have:
\eqabegin
\delta_Q b(z)=\delta_Q \Big(2\beta (z)\partial c(z)+\partial\beta (z)c(z)
\Big)=
\epsilon_Q T(z)
\eqaend
So, comparing with the TCA (15--23), we find two
different candidates for the $G(z)$ current, namely, $G_s (z)$
 and $G_v (z)$ as given in (14). Let us first
consider $G_s (z)$ as our candidate current. This current appeared in the
formulation of Verlinde and Verlinde [6,7]. We note from
(22) that the pole order 2 of
the OPE with $Q(z)$ and
$G(w)$ will generate the $U(1)$ current and it has the form:
\begineq
J(z)\,=\, :c(z)b(z): +2:\beta (z)\gamma (z): .
\endeq

One can easily verify that the currents $T(z)$ in (12), $G_s(z)$ in
(14), $Q(z)$ in (26) and $J(z)$ in (28) indeed satisfy a TCA (15--23)
with central charge $c^{N=2}=-9$. This algebra can be shown to be
isomorphic with the symmetry algebra discussed by Verlinde and Verlinde,
since
the generators in both formulations are related by a similarity
transformation as follows:
\eqabegin
T(z) &=& UT(z)U^{-1} \nn\\
Q_s(z) &=& UQ(z)U^{-1} \equiv U\Big( Q_v (z)+Q_s (z)\Big) U^{-1}\\
G_s (z) &=& UG_s (z) U^{-1}\nn\\
J(z) &=& UJ(z) U^{-1}\nn
\eqaend
where we have defined
\eqabegin
Q_v (z) &=& 2c(z):\beta (z)\partial\gamma (z):
+c(z):\partial\beta (z)\gamma (z):
-:c(z)b(z)\partial c(z):\nn\\
& &\qquad\qquad -2\partial \Big(:\beta(z)\gamma (z): c(z)\Big)-
\frac{3}{2}\partial^{2}c(z)\nn\\
Q_s (z) &=& b(z)\gamma (z)
\eqaend
and $U=:exp[-\frac{1}{2}\oint dz c(z)G_s (z)]:$ is a unitary operator. It is
evident from (29) that the action (1) is individually invariant under $Q_s$
and $Q_v$. This, in fact, helps us to recover another distinct set of
generators which form a TCA with the same central charge $c^{N=2}=-9$.
This second set consists of $T(z)$ and $J(z)$ as given in (12) and (28),
but the fermionic currents have the form:
\eqabegin
Q_v (z) &=& 2c(z):\beta (z)\partial\gamma (z):+
c(z):\partial\beta (z)\gamma (z):-:c(z)b(z)\partial c(z):\nn\\
& &\qquad\qquad -2\partial \Big(:\beta(z)\gamma (z): c(z)\Big)
-\frac{3}{2}\partial^{2} c(z)\\
G_v (z) &=& b(z).
\eqaend
Previous analysis, therefore, brings out two distinct $N=2$ superconformal
structures in the topological gravity formulation of LPW. At this point,
it is natural to ask whether these two $N=2$ structures are parts of a
bigger algebra present in $2d$ topological gravity. We find the answer in
the positive. With the introduction of two more additional currents, we
find that the following eight generators:
\eqabegin
T(z)&=& 2:\beta (z)\partial\gamma (z):+:\partial\beta (z)\gamma (z):-
 2:b(z)\partial c(z):-:\partial b(z) c(z):\nn\\
Q_s (z)&\equiv &G_{1}^{+}(z)=:b(z)\gamma (z):\nn\\
G_s (z)&\equiv &G_{1}^{-}(z)=2\beta (z)\partial c(z)+\partial\beta (z)
 c(z)\nn\\
Q_v (z)&\equiv &G_{2}^{+}(z)=2c(z):\beta (z)\partial\gamma (z):
+c(z):\partial\beta (z)\gamma (z):-:c(z)b(z)\partial c(z):\nn\\
& & -2\partial \Big(:\beta(z)\gamma (z): c(z)\Big)-\frac{3}{2}
\partial^{2}c(z)\\
G_v (z)&\equiv &G_{2}^{-} (z)=b(z)\nn\\
J(z)&=& :c(z)b(z):+2:\beta (z)\gamma (z):\nn\\
J^{++}(z)&=&:b(z)c(z):\gamma (z)-:\gamma^{2} (z)\beta(z):-\frac{3}{2}
\partial\gamma (z)\nn\\
J^{--} (z)&=&\beta (z)\nn
\eqaend
form a twisted ``small'' $N=4$ superconformal algebra. Here we have made
changes of notation according to the $U(1)$ charge. For example,
$G^{\pm} (z)$ have $U(1)$ charges $\pm 1$ and $J^{\pm\pm} (z)$ have
$\pm 2$. It is easy to check that all the $N=4$ generators in (33) are
in fact
currents associated to symmetries of the action (1). The rest of the
relevant OPEs among the currents are given below:
\eqabegin
G^{+}_1 (z)G^{+}_2 (w)&\sim & \frac{2J^{++}(w)}{(z-w)^2} +
 \frac{\partial J^{++}(w)}{(z-w)}\nn\\
G^{-}_1 (z)G^{-}_2 (w)&\sim &-\frac{2J^{--}(w)}{(z-w)^2} -
 \frac{\partial J^{--}(w)}{(z-w)}\nn\\
G^{+}_1 (z)G^{-}_2 (w)&\sim &G^{-}_1 (z)G^{+}_2 (w)\,\,\sim\,\, 0\nn\\
J^{--} (z)J^{++} (w)&\sim &-\frac{3/2}{(z-w)^2} -\frac{J(w)}{(z-w)}\nn\\
G^{+}_1 (z)J^{--}(w)&\sim &-\frac{G^{-}_2 (w)}{(z-w)}\nn\\
G^{+}_2 (z)J^{--}(w)&\sim &\frac{G^{-}_1 (w)}{(z-w)}\\
G^{-}_1 (z)J^{++}(w)&\sim &\frac{G^{+}_2 (w)}{(z-w)}\nn\\
G^{-}_2 (z)J^{++}(w)&\sim &-\frac{G^{+}_1 (w)}{(z-w)}\nn\\
J^{++}(z)J^{++}(w)&\sim &J^{--}(z)J^{--}(w)\,\,\sim\,\, 0\nn\\
G^{+}_1 (z)J^{++}(w)&\sim &G^{+}_2 (z)J^{++}(w)\,\,\sim\,\, 0\nn\\
G^{-}_1 (z)J^{--}(w)&\sim &G^{-}_2 (z)J^{--} (w)\,\,\sim\,\, 0\nn
\eqaend

Note that the currents $J(z)$, $J^{++}(z)$ and $J^{--}(z)$ form an
$SL(2,R)$ Kac-Moody algebra at level $-3/2$. Thus, we found a bigger
symmetry algebra, namely, a twisted ``small'' $N=4$ superconformal
algebra in the topological gravity of LPW (see Figure 1). A similar
twisted $N=4$ superconformal structure have recently been observed in the
formulation of Verlinde and Verlinde in ref. [19].

\begin{figure}
\begin{center}

\begin{picture}(0,0)%
\includegraphics{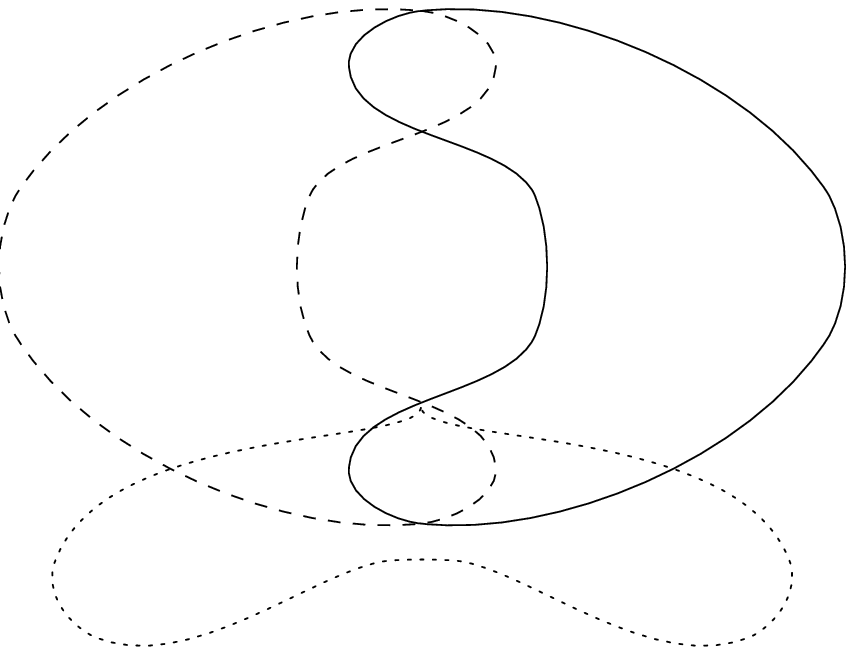}%
\end{picture}%
\setlength{\unitlength}{0.012500in}%
\begingroup\makeatletter\ifx\SetFigFont\undefined
\def\x#1#2#3#4#5#6#7\relax{\def\x{#1#2#3#4#5#6}}%
\expandafter\x\fmtname xxxxxx\relax \def\y{splain}%
\ifx\x\y   
\gdef\SetFigFont#1#2#3{%
  \ifnum #1<17\tiny\else \ifnum #1<20\small\else
  \ifnum #1<24\normalsize\else \ifnum #1<29\large\else
  \ifnum #1<34\Large\else \ifnum #1<41\LARGE\else
     \huge\fi\fi\fi\fi\fi\fi
  \csname #3\endcsname}%
\else
\gdef\SetFigFont#1#2#3{\begingroup
  \count@#1\relax \ifnum 25<\count@\count@25\fi
  \def\x{\endgroup\@setsize\SetFigFont{#2pt}}%
  \expandafter\x
    \csname \romannumeral\the\count@ pt\expandafter\endcsname
    \csname @\romannumeral\the\count@ pt\endcsname
  \csname #3\endcsname}%
\fi
\fi\endgroup
\begin{picture}(271,210)(124,380)
\put(139,523){\makebox(0,0)[lb]{\smash{\SetFigFont{7}{8.4}{rm}$Q_s (z)=G^{+}_1
(z)$}}}
\put(139,487){\makebox(0,0)[lb]{\smash{\SetFigFont{7}{8.4}{rm}$G_s (z)=G^{-}_1
(z)$}}}
\put(306,523){\makebox(0,0)[lb]{\smash{\SetFigFont{7}{8.4}{rm}$Q_v (z)=G^{+}_2
(z)$}}}
\put(309,487){\makebox(0,0)[lb]{\smash{\SetFigFont{7}{8.4}{rm}$G_v (z)=G^{-}_2
(z)$}}}
\put(251,437){\makebox(0,0)[lb]{\smash{\SetFigFont{7}{8.4}{rm}$J(z)$}}}
\put(251,567){\makebox(0,0)[lb]{\smash{\SetFigFont{7}{8.4}{rm}$T(z)$}}}
\put(156,408){\makebox(0,0)[lb]{\smash{\SetFigFont{7}{8.4}{rm}$J^{++} (z)$}}}
\put(312,408){\makebox(0,0)[lb]{\smash{\SetFigFont{7}{8.4}{rm}$J^{--} (z)$}}}
\end{picture}

\end{center}
\caption{\small{Here we depict the different subalgebras
present in the twisted
 ``small" $N=4$ algebra found in $2d$ Topological Gravity. The dashed lines
contains the twisted
$N=2$ subalgebra isomorphic to the symmetry algebra discussed before by the
 Verlindes. Inside the continuous line we present the twisted
 $N=2$ algebra which, through bosonization, have been shown to correspond
to the
$c_M\leq 1$ string theory. Finally, the subalgebra within the dotted line
corresponds to an $SL(2,R)$ Kac-Moody algebra of level $-\frac{3}{2}$.}}
\end{figure}

It should be remarked here that although we find two twisted $N=2$ structures
in $2d$ topological gravity, the currents appeared in the second set (31)
and (32) do not have a natural interpretation in terms of the fields
$\beta (z)$, $\gamma (z)$, $b(z)$ and $c(z)$. A natural interpretation
can be given if we bosonize the deformation ghost of ghost system as follows:
\eqabegin
\beta (z)&=& :\frac{1}{2}\left[\left(\lambda -\frac{1}{\lambda}\right)\partial
\phi_L (z)+i\left(\lambda +\frac{1}{\lambda}\right)\partial\phi_M (z)\right]
e^{-i\lambda \Big(\phi_M (z)-i\phi_L (z)\Big)}:\nn\\
\gamma (z)&=& :e^{i\lambda \Big(\phi_M (z)-i\phi_L (z)\Big)}:
\eqaend
where $\phi_M (z)$ and $\phi_L (z)$ are two bosonic fields and $\lambda$
is a constant. With this bosonization (31) becomes precisely the BRST
current of $c_M\leq 1$  string theory once we identify $\phi_L (z)$ and
$\phi_M (z)$ as the Liouville and matter field of $(p,q)$ minimal models
coupled to gravity system and $\lambda =\sqrt{q/2p}$. This bosonization
have been used in [10] in order to describe topological gravity structure
in any $(p,q)$ minimal model coupled to gravity.

Using this bosonization, we write down all the eight generators (33) in
terms of the matter, Liouville and the reparametrization ghosts $\Big(
b(z),c(z)\Big)$ of $c_M\leq 1$ string theory
as follows:
\eqabegin
T(z)&=& -\frac{1}{2} :\Big(\partial\phi_M (z)\Big)^2: +i Q_M \partial^2
\phi_M (z)-
 \frac{1}{2} :\Big(\partial\phi_L (z)\Big)^2: +i Q_L
\partial^2\phi_L (z)\nn\\
& & - 2:b(z)\partial c(z):-:\partial b(z) c(z):\nn\\
G^{+}_1 (z)&= &:b(z)e^{i\lambda \Big(\phi_M (z)-i\phi_L (z)\Big)}:\nn\\
G^{-}_1 (z)&=& :\left[ \left(\lambda -\frac{1}{\lambda}\right)\left(
\partial\phi_L (z)\partial c(z)+\frac{1}{2}\partial^2 \phi_L (z)c(z)-
\frac{i\lambda}{2}\partial\phi_M (z)\partial\phi_L (z)c(z)\right.\right. \nn\\
& & \left.-\frac{\lambda}{2}\Big(\partial\phi_L (z)\Big)^2 c(z)\right)
 +i\left(\lambda +\frac{1}{\lambda}\right)
\left(\partial\phi_M (z)\partial c(z)+\frac{1}{2}\partial^2 \phi_M (z)c(z)
\right.\nn\\
& & \left.\left. -\frac{i\lambda}{2}\Big(\partial\phi_M (z)\Big)^2 c(z)-
 \frac{\lambda}{2}\partial\phi_L (z)\partial\phi_M (z)c(z)\right)\right]
 e^{-i\lambda \Big(\phi_M (z)-i\phi_L (z)\Big)}:\\
G^{+}_2 (z)&=& :c(z)\left[ -\frac{1}{2}\Big(\partial\phi_M (z)\Big)^2 +i Q_M
  \partial^2\phi_M (z)-\frac{1}{2} \Big(\partial\phi_L (z)\Big)^2 +
 i Q_L \partial^2\phi_L (z)- 2b(z)\partial c(z)\right. \nn\\
& &\left. -\partial b(z) c(z)\right]:
 +\partial\left[ \left( \lambda +\frac{1}{\lambda}\right) c(z)\partial
\phi_L (z)+i\left( \lambda -\frac{1}{\lambda}\right) c(z)\partial\phi_M (z)
\right] -\frac{3}{2}\partial^2 c(z)\nn\\
G^{-}_2 (z)&=&b(z)\nn\\
J(z)&=& :c(z)b(z):-\left( \lambda +\frac{1}{\lambda}\right) \partial
\phi_L (z)-i\left( \lambda -\frac{1}{\lambda}\right) \partial\phi_M (z)\nn\\
J^{++}(z)&=&:\left[ b(z)c(z)-\frac{1}{2\lambda}\Big(i\partial\phi_M (z)-
\partial\phi_L (z)\Big)\right] e^{i\lambda \Big(\phi_M (z)-i\phi_L (z)
\Big)}:\nn\\
J^{--}(z)&=& :\frac{1}{2}\left[ \left(\lambda -\frac{1}{\lambda}\right)
\partial\phi_L (z)+i\left( \lambda +\frac{1}{\lambda}\right)\partial
\phi_M (z)\right] e^{-i\lambda \Big(\phi_M (z)-i\phi_L (z)\Big)}:\nn
\eqaend
\vskip .25cm
Here $Q_L =i\left(\lambda +\frac{1}{2\lambda}\right)$ and
$Q_M =-\left(\lambda -\frac{1}{2\lambda}\right)$ are the background
charges of
 Liouville and matter sectors respectively. Also we note that $J^{++} (z)$
is precisely one of the ground ring generators of $c_M\leq 1$ string theory.
One can explicitly verify that the generators (36) form a twisted ``small''
$N=4$ algebra with central charge $-9$, part of which $\Big(T(z),\,
\,G^{+}_2 (z),\, G^{-}_2 (z)\, {\rm and}\, J(z)\Big)$ has been discussed
in ref. [12,13].

In ref. [13], a one parameter family of twisted $N=2$ superconformal
algebra has been observed in $c_M\leq 1$ string theory. We here note that
for a particular value of this parameter $a_3 =-3/2$, the symmetry algebra
gets enlarged and becomes a part of the twisted $N=4$ superconformal
algebra. Thus, $N=4$ algebra provides a unifying framework for $2d$
topological gravity and $c_M\leq 1$ string theory.

Once we have a twisted $N=4$ superconformal algebra generated by
$T(z)$, $G^{+}_1 (z)$, $G^{-}_1 (z)$, $G^{+}_2 (z)$, $G^{-}_2 (z)$,
$J(z)$, $J^{++}(z)$ and $J^{--}(z)$, we can find another independent
set of generators by applying the so-called ``mirror''
transformation as follows [20]:
\eqabegin
T^{*} (z) = T(z)-\partial J(z),&\qquad\qquad  J^{*} (z)=-J(z)\nn\\
G^{+*}_1 (z) = G^{-}_1 (z),&\qquad\qquad G^{+*}_2 (z)= G^{-}_2 (z)\\
G^{-*}_1 (z) = G^{+}_1 (z), &\qquad\qquad G^{-*}_2 (z)= G^{+}_2 (z)\nn\\
J^{++*}(z)=J^{--}(z),&\qquad\qquad J^{--*} (z)=J^{++}(z)\nn
\eqaend

Written explicitly in terms of the fields of topological gravity, the $N=4$
generators in the mirror representation take the form:
\eqabegin
T^{*}(z)&=&-:\partial\beta (z)\gamma (z):-:b(z)\partial c(z):\nn\\
G^{+*}_1 (z)&=& 2\beta (z)\partial c(z) +\partial\beta (z)c(z)\nn\\
G^{+*}_2 (z)&=&b(z)\nn\\
G^{-*}_1 (z)&=&b(z)\gamma (z)\\
G^{-*}_2 (z)&=&2:\beta (z)\gamma (z):\partial
c(z)-:\partial\beta (z)\gamma (z):c(z)-:c(z)b(z)\partial c(z):-\frac{3}{2}
\partial^2 c(z)\nn\\
J^{*} (z)&=&-:c(z)b(z):-2:\beta (z)\gamma (z):\nn\\
J^{++*} (z)&=&\beta (z)\nn\\
J^{--*} (z)&=&:b(z)c(z):\gamma (z)-:\gamma^2 (z)\beta
(z):-\frac{3}{2}\partial\gamma (z)\nn
\eqaend
It should be pointed out that the generators (38) have precisely the same
form as the $N=4$ generators of $2d$ topological gravity (33) except
$T^{*} (z)$, which makes the physical interpretation of the theory
completely different.
We note that with respect to $T^{*} (z)$, $\beta (z)$, $\gamma (z)$,
$b(z)$ and $c(z)$ have conformal weights $0,1,1,$ and 0 respectively
instead of
$2,-1,2,$ and $-1$ as in the topological gravity.
Identifying $\beta (z)\equiv x(z)$,
$\gamma (z)\equiv\partial\overline{x} (z)$, $b(z)\equiv B(z)$ and $c(z)\equiv
C(z)$, we note that $T^{*}(z)$ in (38) becomes the energy-momentum tensor of a
constrained topological sigma model in complex dimension $d=1$, where $x(z)$
denotes one of the complex target space coordinates of the
topological sigma model
and $\overline{x} (z)$ denotes the holomorphic part of the complex conjugate
of $x(z)$. $B(z)$ and $C(z)$ are the fermionic fields. With the above
identifications, the generators in the ``mirror'' theory takes the form:
\eqabegin
T^{*} (z)&=&-:\partial x(z)\partial\overline{x} (z):-:B(z)\partial C(z):\nn\\
G^{+*}_1 (z)&=&2x(z)\partial C(z)+\partial x(z)C(z)\nn\\
G^{+*}_2 (z)&=&B(z)\nn\\
G^{-*}_1 (z)&=&B(z)\partial\overline{x} (z)\\
G^{-*}_2 (z)&=&2:x(z)\partial\overline{x} (z):\partial C(z)- C(z):\partial x(z)
 \partial\overline{x} (z):-:C(z)B(z)\partial C(z):-\frac{3}{2}
 \partial^2 C(z)\nn\\
J^{*} (z)&=&-:C(z)B(z):-2:x(z)\partial\overline{x} (z):\nn\\
J^{++*} (z)&=&x(z)\nn\\
J^{--*} (z)&=&:B(z)C(z):\partial\overline{x} (z)-
 :x(z)\Big(\partial\overline{x} (z)\Big)^2:-\frac{3}{2}
\partial^2\overline{x} (z)\nn
\eqaend

In this way, we also uncover a twisted $N=4$ structure in the constrained
topological sigma model in complex dimension one. This constrained topological
sigma model has also been shown in [15] to be related with the $c_M\leq 1$
string through the following bosonization:
\eqabegin
x(z)&=&:\left[ b(z)c(z)-\frac{i}{2\lambda}\Big(\partial\phi_M (z)+i\partial
 \phi_L (z)\Big)\right] e^{i\lambda \Big(\phi_M (z)-i\phi_L (z)\Big)}:\nn\\
\partial\overline{x} (z)&=& :e^{-i\lambda \Big(\phi_M (z)-i\phi_L (z)
\Big)}:\\
B(z)&=&:b(z) e^{i\lambda \Big(\phi_M (z)-i\phi_L (z)\Big)}:\nn\\
C(z)&=&:c(z) e^{-i\lambda \Big(\phi_M (z)-i\phi_L (z)\Big)}:\nn
\eqaend
Substituting (40) into (39) one recovers easily the $N=4$ generators of the
$c_M\leq 1$ string theory (36) with $G^{\pm}_1 (z)$ and $G^{\pm}_2 (z)$
interchanged.

To conclude, we have shown that $2d$ topological gravity as formulated
by Labastida, Pernici and Witten possesses a bigger symmetry in the
conformal gauge than what was known before.
We have explicitly shown that it contains
two separate twisted $N=2$ structures one of which is isomorphic to the
symmetry algebra discussed before by Verlinde and Verlinde and the other,
although
does not have a natural interpretation in the topological gravity, can be
related through bosonization to the $N=2$ structure discussed in the
context of $c_M\leq 1$ string theory in ref. [12,13]. These algebras
are then found to be part of a larger algebra, namely they form a twisted
``small'' $N=4$ algebra with the addition of two more generators. Topological
gravity in fact is invariant under the full twisted $N=4$ symmetry.
We thus find
a twisted $N=4$ structure also in $c_M \leq 1$ string theory.
Finally, we studied
the ``mirror'' of this twisted $N=4$ algebra. By using another bosonization
we find that the ``mirror'' theory corresponds to the constrained topological
sigma model in one complex dimension. It would be interesting to investigate
whether a similar extended $N=4$ superconformal structure exists in
$\hat{c}_M\leq 1$ fermionic string theory as well as in W-string theory.

\parbigskip
\parbigskipn
{\bf ACKNOWLEDGEMENTS:}
\parmedskip
We would like to thank A. Ach\' ucarro, J. M. F. Labastida and A. V. Ramallo
for discussions at
various stages of the work. P. M. Ll. would also like to thank the
Departamento de F\'\i sica de Part\'\i culas Elementales
of the Universidade de Santiago de Compostela (Spain),
where this project started, for their hospitality.
The work of P. M. Ll. is supported by the ``Human Capital and Mobility
Program'' of the European Community, that of S. R. was performed as
part of the research program of the ``Stichting voor Fundamenteel Onderzoek
der Materie''(FOM) and that J. M. S. de S. is partially supported by the
DGICYT (PB93-0344).

\parbigskip
\parbigskipn
\noindent {\bf REFERENCES:}
\parmedskip
\begin{enumerate}
\item E. Witten, \np 340 (1990) 281.
\item E. Witten, \cmp 118 (1988) 411.
\item J. M. F. Labastida, M. Pernici and E. Witten, \np 310 (1988)
611.
\item D. Montano and J. Sonnenschein, \np 313 (1989) 258.
\item R. C. Myers and V. Periwal, \np 333 (1990) 536.
\item E. Verlinde and H. Verlinde, \np 348 (1991) 457.
\item R. Dijkgraaf, H. Verlinde and E. Verlinde, preprint PUPT-1217,
IASSNS-HEP-90/80.
\item M. Ademollo et al., \pl 62 (1976) 105 and \np 11 (1976) 77.
     T. Eguchi and A. Taormina, \pl 196 (1987) 75.
\item N. Berkovits and C. Vafa, preprint HUTP-94/A018, KCL-TH-94-12.
\item P. M. Llatas and S. Roy, \pl 342 (1995) 66.
\item B. Gato-Rivera and A. Semikhatov, \pl 288 (1992) 295.
\item M. Bershadsky, W. Lerche, D. Nemeschansky and N. Warner, \np
401 (1993) 304.
\item S. Panda and S. Roy, \pl 317 (1993) 533.
\item S. T. Yau (Ed.), ``Essays on Mirror Manifolds''
(International Press, Hong Kong, 1992).
\item P. M. Llatas and S. Roy, preprint UG-9/94, hep-th/9410233 (to appear
in Phys. Lett. B).
\item E. Witten, Comm. Math. Phys. 117 (1988) 353.
\item D. Friedan, E. Martinec and S. Shenker, \np 271 (1986) 93.
\item T. Eguchi and S. K. Yang, Mod. Phys. Lett. A5 (1990) 1693.
\item A. Boresch, K. Landsteiner, W. Lerche and A. Sevrin, preprint
CERN-TH 7370/94, hep-th/9408033.
\item A. V. Ramallo and J. M. Sanchez-de-Santos, in preparation.
\end{enumerate}
\end{document}